\documentclass[
reprint,
superscriptaddress,
preprintnumbers,
nofootinbib,
 amsmath,amssymb,
 aps,
prd,
floatfix,
]{revtex4-2}

\usepackage{graphicx}
\usepackage{dcolumn}
\usepackage{bm}
\usepackage{orcidlink}
\usepackage{subcaption}

\DeclareCaptionJustification{flushboth}{\leftskip=0pt \rightskip=0pt \parfillskip=0pt plus 1fil\relax}
\usepackage{xcolor}
\usepackage{booktabs}
\usepackage{multirow}
\usepackage{makecell}
\usepackage{xspace}
\makeatletter
\AtBeginDocument{\let\selectlanguage\@gobble}
\makeatother

\newcommand{\refeqn}[1]{Equation~\ref{#1}\xspace}
\newcommand{\tplunge}{t_{\mathrm{plunge}}}
\newcommand{\Msun}{M_\odot}

\begin{document}

\preprint{APS/123-QED}

\title{What Fraction of Extreme Mass Ratio Inspirals That Merge After the LISA Mission Ends Are Detectable?}%

\author{Daniel J. Oliver \orcidlink{0000-0002-7374-6925}}
\thanks{Email: oliverda@oregonstate.edu}%
\affiliation{Oregon State University, 1500 SW Jefferson Ave, Corvallis, OR 97331, USA}

\author{Aaron D. Johnson \orcidlink{0000-0002-7445-8423}}
\affiliation{Science and Technology Institute, Universities Space Research Association, Huntsville, AL 35805, USA}
\affiliation{NASA Marshall Space Flight Center, Huntsville, AL 35812, USA}
\affiliation{Theoretical Astrophysics Group, California Institute of Technology, Pasadena, CA 91125, USA}

\author{Jonathan E. Thompson \orcidlink{0000-0002-0419-5517}}
\affiliation{School of Mathematical Sciences and STAG Research Centre,
University of Southampton, Southampton, SO17 1BJ, United Kingdom}

\author{Curt J. Cutler \orcidlink{0000-0002-2080-1468}}
\affiliation{Theoretical Astrophysics Group, California Institute of Technology, Pasadena, CA 91125, USA}
\affiliation{Jet Propulsion Laboratory, California Institute of Technology, 4800 Oak Grove Drive, Pasadena, CA 91109, USA}

\date{\today}

\begin{abstract}
Stellar-mass compact objects spiraling into massive black holes in galactic nuclei (referred to as ``extreme-mass-ratio inspirals'', or EMRIs) will be an important class of sources for the Laser Interferometer Space Antenna (LISA). A stellar-mass compact object can radiate in the millihertz band for years to decades before it plunges into its massive black hole, so a substantial fraction of the EMRIs that LISA can detect will still be inspiraling when the mission ends. Here we explore that sub-population.  We calculate what fraction of detectable EMRIs will plunge after the mission ends, and for those we calculate how they are distributed as a function of various parameters, such as their post-mission lifetime. To do this, we draw from astrophysically motivated populations, backward-integrate fully relativistic Kerr FastEMRIWaveforms trajectories for a given time to plunge, and compute the full LISA response over both the four-year nominal and ten-year extended observing windows. We calculate the detectable fraction as an explicit function of time to plunge at signal-to-noise ratio (SNR) thresholds of $\rho=10$, 20, and 30. We find that post-mission EMRIs are roughly one in four of the LISA-detectable population at $\rho>20$ over the four-year mission, and roughly one in six over the ten-year extended mission. The in-window detectable fraction varies by nearly a factor of six at $\rho>20$ across the massive black hole spins and secondary masses we ran, while the share stays between 25\% and 33\% for the four-year mission and between 13\% and 20\% for the ten-year one across all of them. The search for EMRIs that plunge more than $\sim 1$ year after the mission ends may  
require a different algorithm than the one used for EMRIs that plunge during the mission, and the fact that post-mission detections are a substantial fraction of all detections gives some urgency to developing a detection pipeline suitable for them. 
\end{abstract}

\maketitle

\section{\label{sec:Intro}Introduction}

The evidence suggests that almost all massive galaxies host a massive black hole (MBH) at their center, and the inner parsecs around such a MBH hold a dense cusp of stars and stellar-mass compact objects (COs) \cite{kormendy_inward_1995, hopman_orbital_2005, kormendy_coevolution_2013, amaro-seoane_laser_2017}. Two-body interactions in this cusp can lead to one of the COs being in a highly eccentric orbit around the MBH, with gravitational radiation reaction causing the CO's orbit to inspiral (and the eccentricity to greatly decrease) until the CO plunges into the MBH. When the MBH mass is in the range $\sim 10^5$ to $10^7\ M_{\odot}$, these will be important sources for the space-based Laser Interferometer Space Antenna (LISA), a joint ESA-NASA mission currently scheduled to launch in 2035. These sources are referred to as extreme-mass-ratio inspirals (EMRIs) and are expected to be among the most important LISA sources~\cite{amaro-seoane_laser_2017, colpi_lisa_2024}.

The CO can be a white dwarf, neutron star or black hole (BH), but BHs are expected to dominate the detection rate, mostly for the same reason that BH-BH inspirals dominate the detection rate for the current ground-based gravitational-wave (GW) detectors: the GWs have larger amplitude, and so can be seen out to a greater distance. So in this paper we will focus on EMRIs where the CO is a BH. In addition to the ``classic'', two-body EMRI formation channel~\cite{hopman_orbital_2005, hopman_effect_2006, amaro-seoane_relativistic_2018}, other possible EMRI formation channels have been identified~\cite{pan_wet_2021, pan_formation_2021, derdzinski_situ_2023, lyu_science_2026, mazzolari_extreme_2022, naoz_enhanced_2023}, but in this paper we will restrict attention to inspiral populations based on the two-body channel.

\begin{figure*}[htb!]
    \centering
    \begin{subfigure}[t]{0.49\textwidth}
        \centering
        \includegraphics[width=\textwidth]{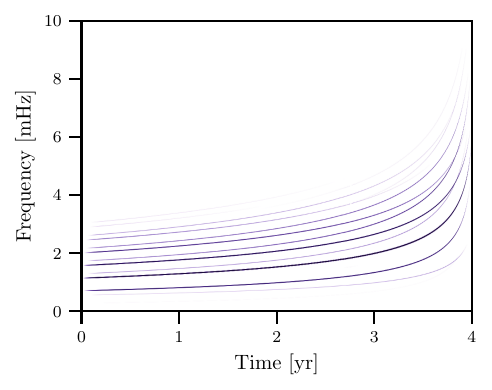}
        \caption{Merges within $4$ years}
        \label{fig:motivation-within}
    \end{subfigure}
    \begin{subfigure}[t]{0.49\textwidth}
        \centering
        \includegraphics[width=\textwidth]{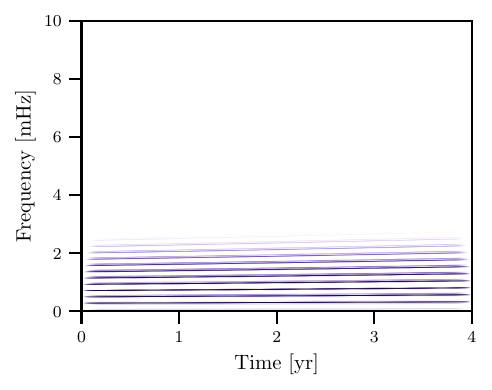}
        \caption{Merges at $20$ years}
        \label{fig:motivation-beyond}
    \end{subfigure}
    \caption{Four-year spectrograms of the strain for one EMRI (with parameters $M=10^6\ M_\odot$, $m=10\ M_\odot$, $a=0.9$, $e_f = 0.05$, at redshift $z=1$), but starting at two different times.  In (a) one sees the waveform starting four years before plunge. In (b), it starts $20$ years before plunge. Clearly, both four-year segments are basically in-band for LISA, but are qualitatively different.}
    \label{fig:tf-motivation}
\end{figure*}

The 2017 LISA proposal specified a nominal science phase of four years \cite{amaro-seoane_laser_2017}, which the LISA Definition Study Report revises to a 4.5-year phase at $\sim82\%$ uptime, resulting in $\sim3.69$ years of accumulated data, with a possible extension out to ten years \cite{colpi_lisa_2024}.  In rough accordance, we adopt four-year and ten-year mission durations throughout, and neglect mission interruptions.

In this paper we divide EMRIs into two categories: the ones that merge during the LISA mission, and the ones that merge after the mission has ended. Figure~\ref{fig:tf-motivation} shows the same EMRI in a four-year window at two different times before plunge. In the first it presents as an unmistakable chirp near the end of the mission, while in the second it sits in band as a slowly evolving, narrowband set of harmonics.

The fact that some detectable EMRIs will merge post-mission has certainly been noted in the literature.  Babak et al. (2017) note that such post-mission EMRIs could be detectable if they are close, but do not include them in their rate estimates~\cite{babak_science_2017}.  Chapman-Bird et al. (2023) do keep post-mission EMRIs in the population they use to train a machine learning selection function~\cite{chapman-bird_rapid_2023}, but do not attempt to address the main questions we pose in this paper. 

Amaro-Seoane et al. (2022) calculate how the total number $N_{\mathrm{det}}$ of EMRI detections (including the post-mission ones) varies with mission duration, $T$ (they find that $N_{\mathrm{det}}$ scales roughly as $T^{3/2}$), but do not break the aggregate count into during-mission and post-mission plunges \cite{amaro_seoane_effect_2022}. And Singh et al. (2026) fold signal-to-noise ratio (SNR)-based selection effects into hierarchical population inference, but do not look specifically at what the post-mission population contributes to the catalog \cite{singh_constraints_2026}. 

By contrast, here the post-mission detections are the main focus. For every source in a set of astrophysically motivated populations, we generate a fully relativistic \texttt{FastEMRIWaveforms} (FEW) \cite{chua_rapid_2021, katz_fast_2021, speri_fast_2024, chapman-bird_efficient_2025} Kerr equatorial-eccentric waveform, propagate it through the full LISA time-delay interferometry response using \texttt{fastlisaresponse} \cite{katz_assessing_2022}, and compute the SNR accumulated over a fixed observing window using \texttt{lisaanalysistools} \cite{lisaanalysistools}. We then characterize the post-mission population in four ways. We measure the detectable fraction as an explicit function of time to plunge at SNR thresholds of 10, 20, and 30. We integrate that curve into the share of the LISA-detectable population that plunges after the mission over both the four-year nominal and the ten-year extended windows. We also study how that share varies with the MBH spin and the mass range of the CO.

The remainder of this paper is organized as follows. Section~\ref{sec:population} describes the population models we draw from. Section~\ref{sec:waveforms} describes the waveform model and the backward-integrated trajectories. Section~\ref{sec:lisaresponse} sets out the LISA response, the noise model, and the two detectability metrics we build from the SNR. Section~\ref{sec:Results} presents our results, taking the detectable fraction, the far tail, the post-mission share, the spin dependence, and the secondary mass dependence in turn. We discuss the implications in Section~\ref{sec:discussion} and conclude in Section~\ref{sec:conclusion}.

Throughout this paper we work in geometric units with $G=c=1$.  Usually that means working in units of seconds, but for familiarity we will often express quantities in units of solar masses, $M_{\odot}$, where the conversion factor is $1\ M_{\odot} = 4.925 \times 10^{-6}\, \mathrm{s}$.

We write the MBH mass as $M$ and the CO mass as $m$ (both in the source frame), the semi-latus rectum as $p$, the eccentricity as $e$, the eccentricity 
at plunge as $e_f$, the time to plunge as $\tplunge$, the SNR as $\rho$, the gravitational-wave frequency as $f$, the detectable fraction as $\mathcal{F}$, and the post-mission share as $\mathcal{S}$. The effective aligned MBH spin (with ``effective'' explained in Section~\ref{sec:waveforms})
is written as $a=a_{\mathrm{max}}\cos\iota$.

\section{\label{sec:population}Population Models}

\begin{table*}[hbt!]
    \centering
    \caption{EMRI population parameters and distributions used in this study, based on Table I of Chua \& Cutler (2022) \cite{chua_nonlocal_2022}, which is built on the M1 model of Babak et al. (2017) \cite{babak_science_2017}. We draw the MBH mass uniformly over their range and we set the spin and the initial orbit as described in Sections~\ref{sec:population} and~\ref{sec:waveforms}. The three completed runs differ only in the CO mass prior, and each draws its own independent set of 4000 sources. Masses are drawn in the source frame and redshifted for waveform generation, and every source is evaluated at both spin magnitudes and at all 72 nodes of the $\tplunge$ grid.} 
    \begin{tabular}{l l l}
        \toprule
        \textbf{Parameter} & \textbf{Description} & \textbf{Distribution} \\
        \midrule
        $M$              & MBH mass in $\Msun$                          & $\mathcal{U}[3\times10^{5},\, 3\times10^{6}]$ \\
        $m$              & CO mass in $\Msun$                           & $10$ (M1);\ \ $\mathcal{U}[10,\, 30]$ (M1-heavy);\ \ $5$ (M1-light) \\
        $e_f$            & Orbital eccentricity at plunge               & $\mathcal{U}[0,\, 0.2]$ \\
        $\cos\iota$      & Cosine of the orbital inclination            & $\mathcal{U}[-1,\, 1]$ \\
        $a$              & Effective aligned spin of the MBH            & $a_{\max}\cos\iota$,\ $a_{\max} = 0.998,\, 0.8$ \\
        $\cos\theta_S$   & Cosine of the polar sky location angle       & $\mathcal{U}[-1,\, 1]$ \\
        $\phi_S$         & Azimuthal sky location angle                 & $\mathcal{U}[-\pi,\, \pi]$ \\
        $\cos\theta_K$   & Cosine of the polar MBH spin angle           & $\mathcal{U}[-1,\, 1]$ \\
        $\phi_K$         & Azimuthal MBH spin angle                     & $\mathcal{U}[-\pi,\, \pi]$ \\
        $\Phi_0,\gamma_0,\alpha_0$ & Initial orbital phases             & $\mathcal{U}[-\pi,\, \pi]$ \\
        $z$              & Redshift to source                          & \refeqn{eq:zprior},\ $0 < z \le 4.5$ \\
        $\tplunge$       & Time to plunge in yr                         & $[0.08,\, 20]$, at 72 discrete times \\
        $N$              & Sources per discrete value of $\tplunge$                  & $4000$ \\
        \bottomrule
    \end{tabular}
    \label{tab:params}
\end{table*}

We draw from three different source populations, all variants of the M1 model from Babak et al. (2017) \cite{babak_science_2017}. Our first model is the same as M1, except that we take the distribution of the MBH mass to be uniform between $3\times 10^5$ and $3\times10^6\ M_{\odot}$, while the M1 model adopts the weighting  $p(M)\propto M^{-1.3}$. This first variant, like the M1 model itself, adopts a delta-function distribution in $m$: $m = 10\ M_{\odot}$.  Our second variant, which we call M1-heavy, adopts a broader $m$ distribution: $m$ is taken as uniformly distributed between $10$ and $30\ M_{\odot}$, informed by the recent GWTC-4 catalog \cite{gwtc-4_population_2026}. Our third variant, which we call M1-light, takes all $m$
to be $5\ M_{\odot}$.

Why do we only consider variants of the M1 model?  (Babak et al. (2017) \cite{babak_science_2017} consider twelve different population models.) While models M1 through M12 give vastly different overall EMRI rates, some experimentation showed us that the fraction of post-mission detections was fairly model-insensitive.  Given that probably none of these models is accurate, considering three variants of M1 seemed good enough for our purposes.

Apart from the different choices of  secondary mass and our assumption here of a uniform distribution in $M$ (that is, uniform within the quoted range), our parameters follow Table~I of Chua \& Cutler (2022) \cite{chua_nonlocal_2022}, which is itself built upon M1 from Babak et al. (2017) \cite{babak_science_2017}. We use the same inclination, three initial phases, sky location, spin orientation, and the redshift distribution from Chua \& Cutler (2022) \cite{chua_nonlocal_2022}. 

We take our eccentricity range from Babak et al. (2017) \cite{babak_science_2017}, who define their eccentricity at the last stable orbit over a uniform distribution $e_f\sim\mathcal{U}(0, 0.2)$. We choose to fix this range at the separatrix partly for its convenience when combined with our backward-integration strategy.  The choice of a flat prior for $e_f$ (with quite limited range) has been challenged in a recent paper; relaxation-driven distributions carry a substantial high-eccentricity tail \cite{mancieri_eccentricity_2026}. Nevertheless we keep this range for more direct comparison to other studies. Our sky location and spin orientation are isotropic over the sky, and all phases are drawn from $\mathcal{U}(-\pi, \pi)$. Redshifts follow the EMRI source distribution of Chua \& Cutler (2022) \cite{chua_nonlocal_2022}, their Equation~(33), 
\begin{equation}
p(z) \propto \frac{D_L^2(z)\,\dot\sigma(z)}{(1+z)^3 H(z)},
\qquad 0 < z \le 4.5,
\label{eq:zprior}
\end{equation}
where $D_L$ is the luminosity distance, $H$ is the Hubble parameter, and $\dot{\sigma}$ is the EMRI rate per unit proper time per unit comoving volume.  We sample Equation~\ref{eq:zprior} for our redshift distribution, and compute luminosity distances with Planck18 using \texttt{Astropy} \cite{astropy:2013, astropy:2018, astropy:2022}.

\section{\label{sec:waveforms}Waveforms and Trajectories}

\begin{figure}
    \centering
    \includegraphics[width=1.0\linewidth]{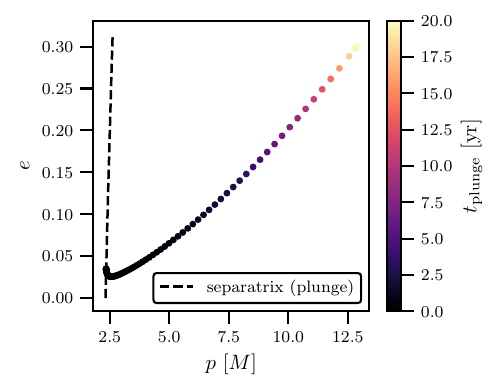}
    \caption{A representative backward-integrated inspiral in the ($p, e$) plane, evolving from its initial orbital elements down to the separatrix (dashed line), with the points colored by time to plunge $\tplunge$. Fixing $\tplunge$ and integrating backward from the separatrix at fixed plunge eccentricity $e_f$ lets us place a source at any distance from merger, and it is the reason $\tplunge$ can serve as the control variable of this study rather than as a derived quantity.}
    \label{fig:trajectory}
\end{figure}

We use the \texttt{FastEMRIWaveforms} Kerr equatorial-eccentric model \cite{chua_rapid_2021, katz_fast_2021, speri_fast_2024, chapman-bird_efficient_2025}, a relativistic, adiabatic, GPU-accelerated waveform family that interpolates fluxes from a pre-computed grid. The flux grid spans dimensionless spin $|a|\leq0.999$, eccentricities $e \leq0.9$, and semi-latus rectum $p\leq200M$ \cite{chapman-bird_efficient_2025}. FEW's speed at computing waveforms makes a study of this size possible.

We chose to use the relativistic FEW model, instead of a FEW ``kludge'' model, because the kludge amplitudes can mis-estimate the SNR by tens of percent over much of our mass range \cite{chapman-bird_efficient_2025,khalvati_impact_2025} and mis-predict horizon redshifts by a similar margin \cite{khalvati_impact_2025}. 

We fix the time to plunge $\tplunge$ and backward-integrate the inspiral from the separatrix to recover the initial $(p_0, e_0)$ for fixed $e_f$ (the eccentricity at plunge). Figure~\ref{fig:trajectory} shows a representative track in the $(p, e)$ plane, with the separatrix denoted as a dashed line, and the individual grid points color coded by time remaining before plunge. We scan $\tplunge$ from 0.08 to 20 years on a grid of 72 time-points, 40 of them spaced evenly from one month to four years and the remaining 32 every half year from 4.5 to 20 years, so that the physically important near-mission-end region is finely resolved while the far tail is still sampled. Each of the $N=4000$ sources drawn per population is integrated back to every time-point, giving $288{,}000$ inspirals per population per spin.

Currently the FEW Kerr model can only handle equatorial orbits, so as a proxy for varying the orbital inclination we instead vary the spin ($a$), setting an effective aligned spin $a=a_{\mathrm{max}}\cos\iota$. The sign of $a$ distinguishes between prograde and retrograde orbits. There is also a practical reason for the proxy, because a population with every source maximally spinning and prograde lies largely outside the current flux grids used by FEW. At maximal spin, backward integration fails much more often due to missing flux values, as the grid is sparse at large eccentricity near the separatrix. With every source fixed at $a=0.998$, the failure rate for backward integration is $\sim67\%$ at $\tplunge=10$ years, and $\sim73\%$ at $\tplunge=20$ years, compared to $1.9\%$ and $2.7\%$ under the proxy.

We have evaluated the robustness of this proxy using FEW's 5PN augmented analytic kludge model, which does allow for inclined orbits \cite{chua_augmented_2017, katz_fast_2021}, and so lets us compare inclined orbits with our proxy version. We find that our proxy mis-estimates the post-mission share of detections by only of order one percentage point.

\section{\label{sec:lisaresponse}LISA Response, Noise, and Detectability}

We propagate every waveform through the full LISA response with \texttt{fastlisaresponse} \cite{katz_assessing_2022}, using first-generation time-delay interferometry (TDI) \cite{armstrong_timedelay_1999} assuming equal arm lengths, in the noise-orthogonal $A$ and $E$ channels \cite{tinto_time-delay_2021}. Every source is sampled at a cadence of $\Delta t = 10$ s and then evaluated against the first-generation sensitivity curves for the $A$ and $E$ channels, so that the response and the noise are taken at the same TDI generation.

The noise model used is SciRDv1~\cite{colpi_lisa_2024, babak_lisa_2021} including the stochastic Galactic confusion foreground from unresolved white dwarf binaries, assembled with the \texttt{SensitivityMatrix} of \texttt{lisaanalysistools}  \cite{lisaanalysistools}. The foreground is modeled as a broken power law, with a knee at the frequency $f_k$ where individual Galactic binaries become resolvable and are removed from the residual \cite{cornish_galactic_2017, robson_construction_2019, babak_lisa_2021, katz_efficient_2025}.
A longer observation time resolves more Galactic binaries, removing these sources from the foreground confusion noise and lowering the knee frequency. The four-year and ten-year windows therefore carry different noise curves, with $f_k=2.09$ mHz at four years against $1.57$ mHz at ten years.

The equal-armlength analytic orbit bundled with \texttt{lisaanalysistools} spans a duration of only five years, so to produce the ten-year mission duration data we regenerate the same analytic orbit to $10.62$ years with \texttt{lisaorbits}  \cite{lisaorbits} (armlength $2.5\times10^9$ m, semi-major axis of $1$~AU with one-day sampling). The longer orbital data reproduces the spacecraft positions of the five-year file to machine precision over the overlapping interval. Given this agreement, we use the ten-year orbit file for both the four-year and ten-year runs, taking just the first four years for the former.

For every EMRI source we compute the optimal SNR, summed over the two noise orthogonal channels,
\begin{equation}\label{eq:snr}
\rho^2 = \sum_{i \in \{A, E\}} 4\int_0^{f_{\mathrm{Nyq}}} \frac{|\tilde{h}_i(f)|^2}{S_i(f)}\,df ,
\end{equation}
where $\tilde{h}_i$ is the frequency-domain TDI strain in channel~$i$, $S_i$ is the corresponding one-sided noise power spectral density including the Galactic foreground, and the integral truncates at the Nyquist frequency $f_{\mathrm{Nyq}}=1/(2\Delta t)=50$ mHz set by our cadence. As a fully coherent matched-filter search is not currently feasible for EMRIs, whose parameter space is far too large for a template bank of the required density, $\rho$ is interpreted as an upper bound on what a realistic semi-coherent search would achieve \cite{chua_nonlocal_2022}. As we see from Equation~\ref{eq:snr}, $\rho^2$ weights the signal power in each frequency bin by the inverse noise power there, so a source is loud when its in-band power accumulates where LISA is most sensitive. A post-mission source is quiet predominantly because its early-inspiral power sits at a lower frequency and lower amplitude than a near-merger EMRI's. For this study, we adopt detection thresholds of $\rho>10, 20,$ and $30$, covering an SNR range used throughout the EMRI literature, and treat the $\rho>20$ threshold that most EMRI rate forecasts adopt as the fiducial threshold \cite{babak_science_2017}.

The primary quantity of interest to this study is the detectable fraction ($\mathcal{F}$) of EMRIs at a given plunge time 
\begin{equation}\label{eq:detfrac}
\mathcal{F}(\tplunge;\rho_{\rm th}) = \frac{N(\rho>\rho_{\rm th}\mid\tplunge)}{N_{\rm valid}(\tplunge)},
\end{equation}
which quantifies the fraction of sources at time $\tplunge$ whose optimal SNR values clear the detection threshold. The denominator counts only the sources whose backward integration spans the full requested $\tplunge$, which is at least $98.2\%$ of all source-node pairs in every population and at both spins, out to $\tplunge=20$ years. Counting those failed integrations as non-detections instead of excluding them shifts $\mathcal{F}$ by at most 0.5 percentage points at any node.

If plunges occur at a constant rate $R$, then the number of systems currently sitting at a time to plunge between $\tplunge$ and $\tplunge+d\tplunge$ is simply $R\ d\tplunge$, so the time to plunge is uniformly distributed and the rate enters only as an overall factor. Weighting Equation~\ref{eq:detfrac} by that flat distribution and normalizing gives the \textit{post-mission share} ($\mathcal{S}$), the fraction of all detections whose merger falls after a mission duration of length $T$,
\begin{equation}\label{eq:share}
\mathcal{S}(T;\rho_{\rm th}) = \frac{\int_T^{20} \mathcal{F}(T)\,d\tplunge}{\int_0^{20} \mathcal{F}(T)\,d\tplunge},
\end{equation}
where $\mathcal{F}(T)$ is the detectable fraction of Equation~\ref{eq:detfrac} evaluated over an observing window of length $T$. Physically, the share is the probability that a randomly chosen detection comes from a source that has not yet merged when observing stops. Because the integral is normalized, $R$ cancels and the share is independent of the absolute EMRI rate, which is highly uncertain.

\section{\label{sec:Results} Results}

We present the results over two observing windows, the four-year nominal mission and the ten-year extended one, with each figure caption giving its own color and linestyles. Table~\ref{tab:results} collects the plateau, post-mission share, and tail numbers quoted below.

\subsection{\label{subsec:detectability} Detectability Versus Time to Plunge}

\begin{figure*}[htb!]
    \centering
    \begin{subfigure}[t]{0.49\textwidth}
        \centering
        \includegraphics[width=\textwidth]{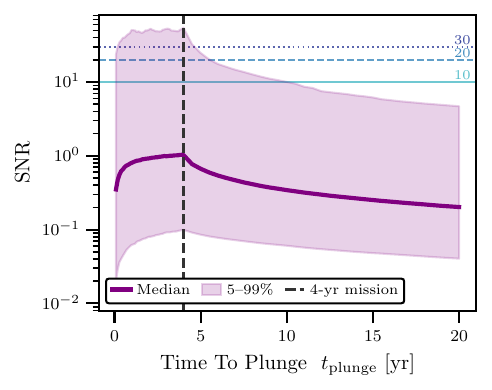}
        \caption{Median and $5-99\%$ TDI-SNR band}
        \label{fig:m1-snr-T4}
    \end{subfigure}
    \begin{subfigure}[t]{0.49\textwidth}
        \centering
        \includegraphics[width=\textwidth]{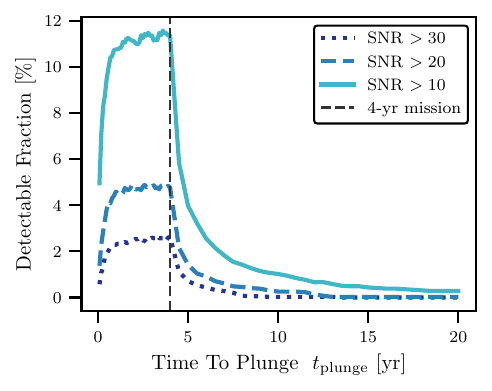}
        \caption{Detectable fraction at $\rho > 10/20/30$}
        \label{fig:m1-detfrac-T4}
        \end{subfigure}
    \caption{Detectability of the fiducial M1 population (4000 sources per node, $m=10\ \Msun, a=0.998\cos\iota$, SciRDv1 noise with the Galactic confusion foreground) over the four-year mission, against time to plunge $\tplunge$. Every SNR is computed from a fully relativistic FEW Kerr equatorial-eccentric waveform propagated through the full LISA response in the A and E channels. (a) shows the median SNR and the band spanning the 5th to 99th percentiles at each node. (b) shows the detectable fraction of Equation~\ref{eq:detfrac} at SNR thresholds of 10, 20, and 30, drawn in teal (solid), blue (dashed), and indigo (dotted). The same three thresholds are marked by horizontal lines in (a), and the vertical dashed line in both panels marks the end of the nominal mission. The detectable fraction plateaus while the sources plunge in band, so detections come from the top of the band rather than the middle of it. The upper edge of the band in (a) is the 99th percentile, so wherever it crosses a threshold line, exactly 1\% of sources lie above that SNR. That is the same $\tplunge$ at which the curve for that threshold in (b) passes through 1\%.}
    \label{fig:m1_snr_detfrac4}
\end{figure*}

We start with the SNR versus time to plunge seen in Figure~\ref{fig:m1_snr_detfrac4} (a). We plot the 5-99\% interval for the post-TDI SNR in shaded purple with its median as a solid line. All curves rise sharply within the first year as EMRI signals need time to build up SNR, after the first year it reaches a plateau for the remainder of the mission duration, and then has a sharp decrease out to $\tplunge=20$ years. In Figure~\ref{fig:m1_snr_detfrac4} (b), we plot the detectable fraction versus time to plunge using Equation~\ref{eq:detfrac} at three SNR detection thresholds. Somewhat expectedly, we see similar behavior as in the SNR versus $\tplunge$ figure, where the detectable fraction has a very sharp increase within the first year as the signals have time to build SNR in band, where it then plateaus until reaching the mission duration, where it then has a steep decline in detectable sources as the merger event is no longer seen by LISA. Interestingly, only the $\rho>30$ curve reaches zero within the grid, at $\tplunge\simeq13.5$ years, while $\rho>20$ flattens onto a floor near 0.03\% and the $\rho>10$ threshold still holds 0.28\% at $\tplunge=20$ years. 

When looking over the entire $\tplunge$ grid, the median SNR is approximately 0.5, while the 99th percentile reaches $\rho\simeq35$. Almost every source is therefore too quiet to detect, and the ones we do count come from the loud tail of the distribution rather than anywhere near the median.

\begin{figure*}[htb!]
    \centering
    \begin{subfigure}[t]{0.49\textwidth}
        \centering
    \includegraphics[width=\textwidth]{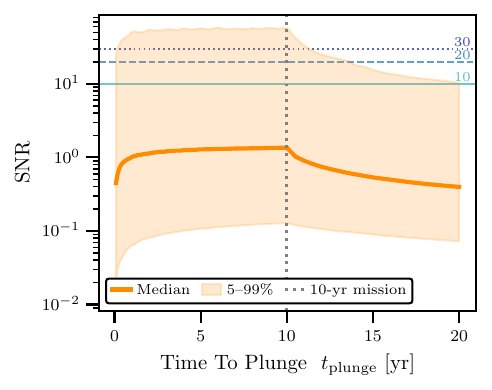}
        \caption{Median and $5-99\%$ TDI-SNR band}
        \label{fig:m1-snr-T10}
    \end{subfigure}
    \begin{subfigure}[t]{0.49\textwidth}
        \centering
        \includegraphics[width=\textwidth]{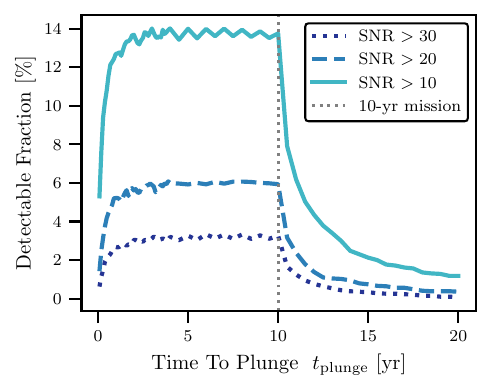}
        \caption{Detectable fraction at $\rho > 10/20/30$}
        \label{fig:m1-detfrac-T10}
    \end{subfigure}
    \caption{Similar to Figure~\ref{fig:m1_snr_detfrac4}, but for the ten-year extended mission, with the vertical dotted line now at ten years. The population and the spin are identical. Only the observing window differs, together with the duration-dependent Galactic confusion foreground that comes with it. The plateau is higher because the longer observation resolves more of the Galactic foreground, and the tail extends further because of the additional in-band time. The detectable fraction curves are visibly more jagged across the plateau than in Figure~\ref{fig:m1_snr_detfrac4}, since the $\tplunge$ grid is sparser between four and ten years than it is between zero and four. The separation from the four-year case is concentrated in $4<\tplunge<10$ years, where the longer mission still catches the loud near-merger phase and the shorter one sees only the early inspiral. The far detectable tail is clearer here than in Figure~\ref{fig:m1_snr_detfrac4}: at $\tplunge=20$ years the $\rho > 10$ and $\rho>20$ curves hold at $1.18\%$ and $0.36\%$ detectable fractions respectively.}
    \label{fig:m1_snr_detfrac10}
\end{figure*}

Figure~\ref{fig:m1_snr_detfrac10} shows the same quantities for the extended mission duration of $T=10$ years. The curves keep their shape but sit higher, largely because the longer observation resolves more of the Galactic confusion foreground and so lowers the effective noise. The detectable tail extends further because the sources spend longer in band, holding $1.18\%$ for the ten-year mission at $\tplunge=20$ years against $0.28\%$ for the four-year mission.

The sources that survive in the detectable tail are the nearby, low redshift ones. Taking those with $\tplunge\geq15$ years and $\rho>10$ detectable threshold, the four-year survivors have a median redshift of $z=0.31$ and none above $z=0.54$, while the ten-year survivors have a median of $z=0.41$ and a maximum of $z=1.12$. Both medians are below the M1 population's own median redshift of $z=2.08$. The reason the detectable tail sources are close by is because the amplitude of the signal scales inversely with distance, so the closest sources are loud enough to cross the detectability threshold even without a merger.

\subsection{\label{subsec:beyondshare} The post-mission Share}

To quantify, we can use Equation~\ref{eq:share} to integrate the detectable fraction curves and determine the detectability of our post-mission EMRIs. We compile this information in Table~\ref{tab:results}, which shows, for each of our three population models, the detectable fraction for all three SNR thresholds over both mission durations for the plateau (in-window), the post-mission share, and the detectable tail (for the $\rho>10$ threshold only). The headline result is that the post-mission share is a non-negligible fraction of the LISA-detectable population, with roughly one in four of the detectable sources being post-mission mergers at $\rho>20$ (27\% for M1, 25\% for M1-heavy, and 31\% for M1-light for the four-year mission). The post-mission share is larger for the shorter mission for two reasons. The first is that in Equation~\ref{eq:share}, moving $T$ from four to ten years narrows the numerator from 16 years of the $\tplunge$ grid to 10 while widening the in-window part of the denominator from four years to ten, which moves the share down. The second is that $\mathcal{F}(T)$ rises everywhere for the longer window because the sources spend longer in band and because the Galactic confusion foreground is more resolved, which rises most in the far tail. The ten-year mission duration shows that for the $\rho>20$ SNR threshold, roughly one in six detectable EMRIs will merge beyond the mission duration (16\% for M1, 13\% for M1-heavy, and 18\% for M1-light). The longer mission, however, detects more sources in absolute terms. 

One interesting aspect that can be gleaned from the results is that the SNR threshold has very little effect on the post-mission share, with M1 having 33\%, 27\%, and 23\% for the $\rho>10, 20,$ and $30$ respectively. The other two models behave similarly. While raising the threshold cuts the detectable fraction sharply at every $\tplunge$, it cuts the in-mission and post-mission parts by similar factors, so the ratio between them survives. Because the plunge rate is assumed to be steady, it enters Equation~\ref{eq:share} as the same constant factor in the numerator and denominator, canceling, so the share does not depend on the absolute EMRI rate.

\begin{table*}[t]
\caption{Summary of the three completed populations at both spin magnitudes, computed directly from the source catalogs with SciRDv1 noise and the Galactic confusion foreground. For each population, spin, and observing window we list three quantities, the first two at SNR thresholds of 10, 20, and 30. The plateau is the in-window detectable fraction, the mean of Equation~\ref{eq:detfrac} over $1 \leq \tplunge \leq T - 0.5$ years. The share is the post-mission fraction of all detections, Equation~\ref{eq:share}, obtained by assuming a steady state in which the time to plunge is uniformly distributed, which makes it independent of the uncertain absolute EMRI rate. The tail is the detectable fraction at $\rho > 10$ evaluated at $\tplunge=20$ years, five times the nominal mission length before merger. Two patterns run through the table. The plateau rises steeply with secondary mass, by more than a factor of three at $\rho>20$ from M1-light to M1-heavy, and falls to 36\% when the spin is lowered. The share doesn't follow either pattern. It stays near one in four over the four-year mission and near one in six over the ten-year one at $\rho>20$, across every population and spin.}
\label{tab:results}
\begin{ruledtabular}
\begin{tabular}{llcc@{\hspace{2.2em}}ccc@{\hspace{2.2em}}ccc@{\hspace{2.2em}}c}
 & & & & \multicolumn{3}{c}{Plateau (in-mission) [\%]} & \multicolumn{3}{c}{Post-mission Share [\%]} & Tail [\%]\\
\cline{5-7}\cline{8-10}
Model & $m\ [\Msun]$ & $a_{\mathrm{max}}$ & Mission & $\rho{>}10$ & $\rho{>}20$ & $\rho{>}30$
 & $\rho{>}10$ & $\rho{>}20$ & $\rho{>}30$ & $\rho{>}10$\\
\colrule
\multirow{4}{*}{M1} & \multirow{4}{*}{$10$} & \multirow{2}{*}{$0.998$} & 4-yr & $11.2$ & $4.7$ & $2.5$ & $33$ & $27$ & $23$ & $0.28$\\
 & & & 10-yr & $13.6$ & $5.8$ & $3.0$ & $19$ & $16$ & $15$ & $1.18$\\
 & & \multirow{2}{*}{$0.8$} & 4-yr & $8.3$ & $3.2$ & $1.8$ & $34$ & $26$ & $20$ & $0.18$\\
 & & & 10-yr & $10.6$ & $4.1$ & $2.2$ & $20$ & $18$ & $15$ & $1.08$\\
\colrule
\multirow{4}{*}{M1-heavy} & \multirow{4}{*}{$\mathcal{U}(10,30)$} & \multirow{2}{*}{$0.998$} & 4-yr & $16.6$ & $8.2$ & $4.7$ & $30$ & $25$ & $24$ & $0.39$\\
 & & & 10-yr & $19.8$ & $10.1$ & $5.9$ & $17$ & $13$ & $12$ & $1.48$\\
 & & \multirow{2}{*}{$0.8$} & 4-yr & $13.4$ & $6.0$ & $3.4$ & $31$ & $26$ & $24$ & $0.40$\\
 & & & 10-yr & $16.6$ & $7.8$ & $4.4$ & $17$ & $14$ & $14$ & $1.21$\\
\colrule
\multirow{4}{*}{M1-light} & \multirow{4}{*}{$5$} & \multirow{2}{*}{$0.998$} & 4-yr & $7.2$ & $2.3$ & $1.0$ & $35$ & $31$ & $28$ & $0.15$\\
 & & & 10-yr & $9.0$ & $2.9$ & $1.2$ & $21$ & $18$ & $19$ & $0.71$\\
 & & \multirow{2}{*}{$0.8$} & 4-yr & $5.1$ & $1.4$ & $0.6$ & $36$ & $33$ & $29$ & $0.15$\\
 & & & 10-yr & $6.8$ & $1.9$ & $0.8$ & $22$ & $20$ & $20$ & $0.58$\\
\end{tabular}
\end{ruledtabular}
\end{table*}

\subsection{\label{subsec:spin} Dependence on the Massive Black Hole Spin}

\begin{figure*}[htb!]
    \centering
    \begin{subfigure}[t]{0.49\textwidth}
        \centering
        \includegraphics[width=\textwidth]{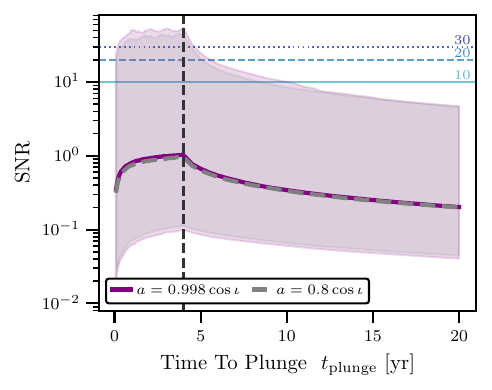}
        \caption{Median and $5-99\%$ TDI-SNR band}
        \label{fig:m1-spin-T4}
    \end{subfigure}
    \begin{subfigure}[t]{0.49\textwidth}
        \centering
        \includegraphics[width=\textwidth]{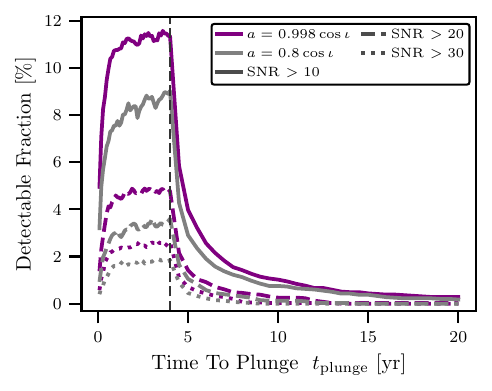}
        \caption{Detectable fraction}
        \label{fig:m1-spin-detfrac-T4}
    \end{subfigure}
    \caption{Spin dependence for the M1 population. The same 4000 sources underlie every curve, so the only thing that changes is the spin magnitude, and the vertical line marks the end of the mission. (a) Median SNR and its 5-99\% band for $a=0.998\cos\iota$ against $a=0.8\cos\iota$, which are hard to tell apart, differing by only about 6\% in the median. (b) The detectable fraction for the same two spins at $\rho>10, 20,$ and $30$, with the spin carried by the colors in (a) and the threshold by the linestyle. The two spins separate clearly across the in-window plateau, by 25\% to 31\% depending on the threshold.}
    \label{fig:m1-spin}
\end{figure*}

To examine the effects of MBH spin on the post-mission share, we examined $a=a_{\mathrm{max}}\cos\iota$ for two different spin values $a_{\mathrm{max}}=0.998$, and $0.8$ on the same 4000 sources and the same $\tplunge$ grid. The results can be seen for the four-year window in Figure~\ref{fig:m1-spin}, where (a) shows the median SNR and (b) the detectable fraction. Here we see that the median SNR changes by a median curve ratio of only $\sim1.056$ for the four-year window and $\sim1.063$ for the ten-year window, which is small enough that the two median curves are hard to tell apart by eye.

The detectable fraction moves a large amount more, as Table~\ref{tab:results} shows. Lowering the spin cuts the in-window plateau by roughly a quarter to a third for M1 at every threshold and in both windows, while the post-mission share moves by no more than 3 percentage points and not always in the same direction. The cut is shallower for M1-heavy, at 16\% to 28\%, and deeper for M1-light, at 25\% to 36\%, but the share is unmoved in all three, because the post-mission part of each curve decreases by nearly the same factor as the in-window part. Spin therefore changes how many EMRIs LISA detects without changing what fraction of them will still be inspiraling when the mission ends.

A 6\% shift in the median SNR produces a 30\% shift in the detectable fraction because the spin changes the shape of the SNR distribution rather than shifting it. That happens because $a=a_{\mathrm{max}}\cos\iota$ ties the spin magnitude to the inclination, so a larger $a_{\mathrm{max}}$ makes the prograde sources louder and the retrograde ones quieter. Splitting the four-year population at $\cos\iota=0$ highlights the effect. At $\rho>20$ the prograde plateau climbs from 5.3\% to 8.4\% between the two spins while the retrograde drops from 1.3\% to 1.1\%, so the prograde gains far more than the retrograde loses. This is a known asymmetry, since raising the spin shrinks the effective size of the MBH for prograde orbits and enlarges it for retrograde ones, and the two effects do not cancel \cite{amaro-seoane_role_2013}. Detections come from the upper tail, which is almost entirely prograde, and those are the sources the spin affects the most. The median SNR sits far below any of the detection thresholds, among the sources the spin hardly impacts. Inclination therefore matters a lot for an individual source, since it sets the effective spin in the first place, and 88\% of the detections at $\rho>20$ come from prograde orbits against 50\% of the parent population. That does not propagate into the share because the population is isotropic in $\cos\iota$, so the prograde gain and the retrograde loss are integrated over together.

\subsection{\label{subsec:mass} Dependence on the Secondary Mass}

\begin{figure*}[htb!]
    \centering
    \begin{subfigure}[t]{0.49\textwidth}
        \centering
        \includegraphics[width=\textwidth]{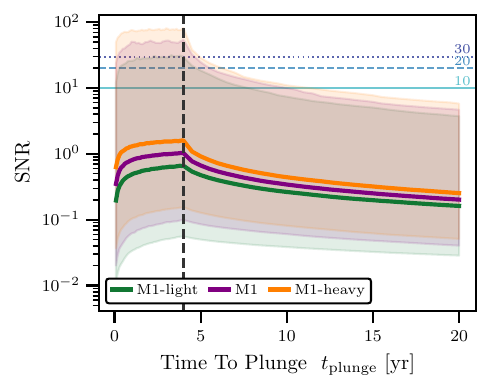}
        \caption{Median and $5-99\%$ TDI-SNR band}
        \label{fig:mass_comp_snr}
    \end{subfigure}
    \begin{subfigure}[t]{0.49\textwidth}
        \centering
        \includegraphics[width=\textwidth]{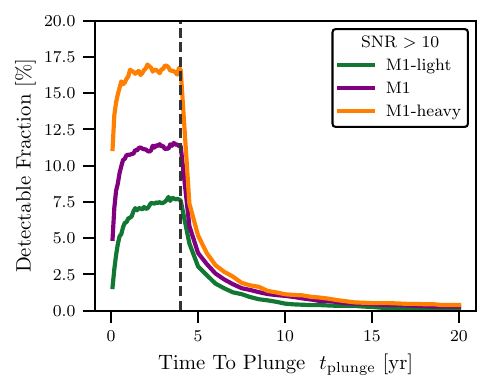}
        \caption{Detectable fraction}
        \label{fig:mass_comp_detfrac}
    \end{subfigure}
    \caption{Secondary-mass sensitivity at $a=0.998\cos\iota$, over the four-year mission with 4000 sources per node. Both panels show M1-light ($m=5\ \Msun$), M1 ($m=10\ \Msun$), and M1-heavy ($m\sim\mathcal{U}(10,30)\ \Msun$) against time to plunge. (a) shows the median SNR and its 5-99\% band for the four-year mission. (b) The detectable fraction at $\rho>10$, the threshold at which all three populations retain a visible far tail, with the ordering the same at $\rho>20$ and $\rho>30$ as seen in Table~\ref{tab:results}.}
    \label{fig:mass_comp}
\end{figure*}

Across the ranges considered, the secondary mass has a far larger impact on the detectability than MBH spin. A heavier secondary radiates more strongly and raises the SNR at every plunge time, so the M1-heavy population ($m\sim\mathcal{U}(10,30)\ \Msun$) sits systematically above the M1 population ($m=10\ \Msun$) in Figure~\ref{fig:mass_comp}, and over the ten-year window as well in Table~\ref{tab:results}, while the M1-light ($m=5\ \Msun$) sits lower than the other two across all $\tplunge$.

The plateau climbs monotonically with secondary mass at every threshold and in both windows, by more than a factor of three from M1-light to M1-heavy at $\rho>20$.

The post-mission shape, however, is preserved across the comparison, and all three populations collapse onto a similar decline after the mission, as Figure~\ref{fig:mass_comp_detfrac} shows. The four-year share at $\rho>20$ runs 31\%, 27\%, and 25\% for M1-light, M1, and M1-heavy, moving in the opposite direction to the plateau and by far less. As with the spin, the plateau depends strongly on the secondary mass while the post-mission share does not. The reason is that a heavier secondary mass raises the SNR of both the in-mission and post-mission sources, but it raises the in-mission ones more, which is why the post-mission share decreases slightly as the secondary mass rises rather than holding exactly still.

\section{\label{sec:discussion}Discussion and Implications}

Three things follow from the detectable fraction curves of Section~\ref{sec:Results}. The first is that post-mission EMRIs are a substantial part of what LISA will detect rather than a marginal population at the edge of the catalog, so a data-analysis plan that assumes every detected EMRI merges in band would be wrong for about a quarter of its sources. The second is that the share is far more stable than the detection count it is drawn from.  The astrophysical choices that move the in-window plateau by nearly a factor of six at $\rho>20$ affect the detectability of in-mission and post-mission sources by roughly the same amount, leaving the share close to where it started. The third is that detectability does not switch off when a source stops merging in band but decays smoothly, leaving a small population drawn from the nearby universe that is still visible up to twenty years from merger. All three still hold for a ten-year mission. LISA detects more sources and the far tail is larger, while the post-mission share falls to about one in six.

The FEW Kerr model is equatorial, so every number we report uses the effective aligned spin $a=a_{\mathrm{max}}\cos\iota$ in place of a true inclination. Checking this against inclined orbits with the augmented analytic kludge model moves the post-mission share by about one percentage point (Section~\ref{sec:waveforms}). The assumption that an equatorial orbit never passes through the transient resonances on an inclined inspiral \cite{lynch_fast_2024, van_de_meent_fast_2018} could shift the post-mission share either way. Additionally, we tested our assumption of a uniform MBH mass rather than the $p(M)\propto M^{-1.3}$ weighting in Babak et al. (2017) \cite{babak_science_2017} or Chua \& Cutler (2022) \cite{chua_nonlocal_2022} by rerunning M1 holding every other source parameter fixed. With this test, the post-mission share moves by only $\sim1$ percentage point.

Three of the assumptions could make the post-mission population larger than we report, the eccentricity prior ($\mathcal{U}(0, 0.2)$) being the most consequential. Examining just the more eccentric half, $e_f \geq 0.1$, shows that it carries a post-mission share 7 percentage points above the lower half $e_f \leq 0.1$ over the four-year window at $\rho>20$. That is a wider spread than the spin produces and is comparable to the effect from secondary mass. Relaxation driven distributions are expected to carry a substantial tail above that range \cite{mancieri_eccentricity_2026}, so the flat prior we keep is the conservative choice. The second assumption is the set of trajectories we drop, the ones whose backward integration fails due to the limited parameter space of the FEW flux grid (at most $\sim2\%$ of the population, as shown in Section~\ref{sec:lisaresponse}). They are not scattered through the parameter space, but lie mostly in the high eccentricity regime at longer times to plunge, which is also where the post-mission sources are the loudest. The reason is that an eccentric orbit spreads its power over many harmonics, some of which land where LISA is most sensitive, while a circular source far from plunge radiates at a single frequency and remains relatively quiet. Dropping these trajectories therefore removes sources that were likely to have been detected. That corner of parameter space is also the one identified as poorly covered by the current flux grids \cite{mancieri_eccentricity_2026}, so the first two assumptions are not independent of one another. The third assumption is simply that we chose to stop the $\tplunge$ grid at twenty years, which truncates the far tail and leaves the share at a lower bound in that respect. 

The assumptions that work the other way come down to a single choice. We count a source as detected when its optimal SNR $\rho$ clears the threshold, and a real search will recover only a part of the SNR. Losing a fixed fraction of $\rho$ is the same as raising the threshold, and the share only moves weakly with the threshold, so a loss of tens of percent costs only a few percentage points of the post-mission share. This is optimistic, because the loss is not a fixed fraction. Semi-coherent detection efficiency already degrades toward the longest plunge times searched \cite{speri_single-harmonic_2026}, which is the direction this population lies in. Three further effects push in the same direction, and we leave them for future work. Gaps in the data cost SNR, and we have treated all of our sources as continuous with stationary noise \cite{colpi_lisa_2024}. Bright Galactic binaries can imitate an EMRI \cite{khukhlaev_assessing_2025}, and we expect a slowly evolving EMRI to be the easiest to imitate. EMRIs have also not yet been folded into a global fit, so what that could cost is currently unknown \cite{khukhlaev_assessing_2025,littenberg_prototype_2023, katz_efficient_2025}.

The steady-state rate behind Equation~\ref{eq:share} is the last of the assumptions and the least worrying, since the timescales that govern EMRI formation are much longer than the LISA mission by many orders of magnitude \cite{hopman_orbital_2005, amaro-seoane_relativistic_2018}. Taken together, the eccentricity prior is the assumption most able to move the answer, and the more eccentric prior that relaxation-driven models predict would raise the post-mission share, while a realistic search, which recovers only part of the optimal SNR, would lower it by a small amount, as far as we can bound it. The one in four we report is therefore a floor rather than a ceiling on the post-mission share. 

This work has a practical consequence for future EMRI search design, since a post-mission EMRI presents as a slowly evolving, narrowband set of harmonics (Figure~\ref{fig:motivation-beyond}) rather than the upward chirp a search built around merging sources is tuned to find. Such sources suit searches that treat the time to plunge as an explicit parameter: the single-harmonic statistic of Speri et al. (2026) \cite{speri_single-harmonic_2026} treats it as an explicit search parameter, and the phase-incoherent search of Strub et al. (2026) \cite{strub_searching_2026} uses it to set the initial orbit in its first stage. Both are currently built for sources that plunge within the observation. A search of that kind must also contend with the non-stationary noise of a years-long observation, where a Wilson basis has advantages over the frequency domain because it localizes in time as well as frequency \cite{cornish_time-frequency_2020, cornish_non-stationary_2025, vajpeyi_explicit_2026,johnson_wdm_2026}.

Post-mission EMRIs also propagate downstream into population inference and into catalog construction. A hierarchical framework defining detectability by an SNR threshold folds the post-mission EMRIs into its selection function whether or not it treats them separately \cite{chapman-bird_rapid_2023,singh_constraints_2026}, and because they are drawn from the loud tail and sit at systematically lower redshift, a selection function calibrated on in-band mergers would likely misrepresent them specifically. In a catalog built from a global fit, a construction demonstrated so far on Galactic binaries and expected to extend to other source types \cite{johnson_petra_2025}, they would be the marginal entries, since a source that never merges in band offers no loud final approach for the fit to lock onto. The sources we do not resolve matter too, since the far more numerous unresolved EMRIs blend into a stochastic confusion background \cite{pozzoli_computation_2023, oliver_gravitational_2024, oliver_gravitational_2026} from which EMRI population parameters may themselves be recoverable. The reason is that EMRIs as a population chirp faster than the Galactic binaries they sit alongside, and so leave a distinct time-frequency imprint \cite{ji_time-frequency_2025}, although we expect that discriminant to be weakest for exactly the slowly evolving sources considered here.

\section{\label{sec:conclusion}Conclusions}

We set out to measure how detectable an EMRI is as a function of time to plunge, and what share of the population LISA can detect will still be inspiraling when observing stops. We did this for three astrophysically motivated populations spanning secondary masses from 5 to 30 solar masses, drawing 4000 sources per population and running each at two spin magnitudes. For every source we fixed the time to plunge by integrating backward a fully relativistic FEW Kerr trajectory from the separatrix, propagated the waveform through the full LISA response, and computed the optimal SNR over both a four-year and a ten-year observing window. 

At our fiducial threshold of $\rho>20$, roughly one in four of the detectable population merges beyond the four-year mission, between 25\% and 33\% across every population and spin we ran, and about one in six beyond the ten-year one. At $\rho>10$ the four-year share rises to roughly a third.  The far tail is thin but real, since at the lowest threshold no population reaches zero detectability anywhere on the grid, leaving sources still visible twenty years before they merge and drawn almost entirely from low redshift. This post-mission population is set by the loud, nearby sources of the SNR distribution rather than by the typical source, which is too quiet to detect. We also report the post-mission share alongside the detectable fraction because the in-window plateau can move by nearly a factor of six at $\rho>20$ across the spins and secondary masses we ran while the post-mission share barely changes. The share is also independent of the uncertain absolute EMRI rate, since it is a ratio of detections.

The clearest way to improve on this work is a generic-inclination waveform model fast enough to run at this scale, which would replace the effective aligned spin with a fully generic one, and remove the largest waveform approximation in this study \cite{hughes_adiabatic_2021}. Extended flux grids reaching to higher eccentricity in the strong field, as called for in Mancieri et al. (2026) \cite{mancieri_eccentricity_2026}, would also close the corner of parameter space where our backward integration fails. Pushing the $\tplunge$ grid past twenty years, out to where the detectable tail finally reaches zero, would cover the rest of this population. Lastly, repeating this study under a relaxation-driven eccentricity prior rather than a flat one would test our most consequential assumption.

The largest remaining question is to determine how much of this population survives a realistic EMRI search rather than just a detection threshold based on optimal SNR. Folding these sources explicitly into a realistic search alongside plunging EMRI signals will provide better insights into what they contribute to the EMRI science case, which remains to be quantified.  What this study does establish is that these sources exist, and that roughly a quarter of what LISA sees over its nominal mission will be systems it never watches merge. 

\section*{Acknowledgments}

DJO's work was supported by NSF Physics Frontiers Center award No. 2607948. DJO's and ADJ's work was supported by NSF Physics Frontiers Center award No. 2020265.
ADJ and CC were supported by the LISA Project Office during this work.
JET acknowledges support from the NASA LISA Preparatory Science grant 20-LPS20-0005. CC's research was carried out at the Jet Propulsion Laboratory, California Institute of Technology, under a contract with the National Aeronautics and Space Administration (80NM0018D0004). Anthropic’s Claude agentic AI (Opus 4.7, Opus 4.8, and Opus 5) was used in the production of
this paper for reviewing the manuscript, and assisting with developing and debugging the code. The authors directed the tool throughout and have reviewed everything produced with agentic AI for correctness and take full responsibility for the content of this paper. This work used the computing resources at the Center for Quantitative Life Sciences \texttt{Wildwood} cluster at Oregon State University.

\textit{Software:} \texttt{Black Hole Perturbation Toolkit} \cite{BHPToolkit}, \texttt{FastEMRIWaveforms} \cite{chua_rapid_2021, katz_fast_2021, speri_fast_2024, chapman-bird_efficient_2025}, \texttt{fastlisaresponse} \cite{katz_assessing_2022}, \texttt{lisaanalysistools} \cite{lisaanalysistools}, \texttt{lisaorbits} \cite{lisaorbits}, \texttt{Matplotlib} \cite{matplotlib}, \texttt{NumPy} \cite{numpy}, \texttt{SciPy} \cite{scipy}, \texttt{AstroPy} \cite{astropy:2013,astropy:2018,astropy:2022}.\par

\clearpage

\bibliography{post-mission_emri}

\end{document}